\documentclass[11pt]{asaproc}

\usepackage{graphicx}
\usepackage{amsmath}
\usepackage{xurl}
\usepackage{natbib}

\usepackage{times}

\title{ Forecasting Hierarchical Time Series}

\author{Seema Sangari\thanks{School of Data Science and Analytics, Kennesaw State University, 3391 Town Point Dr. NW,Kennesaw, GA 30144} \and  Xinyan Zhang\thanks{School of Data Science and Analytics, Kennesaw State University, 3391 Town Point Dr. NW,Kennesaw, GA 30144} }
\begin{document}

\maketitle

\begin{abstract}
This paper addresses a common problem with hierarchical time series. Time series analysis demands the series for a model to be the sum of multiple series at corresponding sub-levels. Hierarchical Time Series presents a two-fold problem. First, each individual time series model at each level in the hierarchy must be estimated separately. Second, those models must maintain their hierarchical structure over the specified period of time, which is complicated by performance degradation of the higher-level models in the hierarchy.  This performance loss is attributable to the summation of the bottom-level time series models. In this paper, the proposed methodology works to correct this degradation of performance through a top-down approach using odds, time series and systems of linear equations. Vertically, the total counts of corresponding series at each sub-level are captured while horizontally odds are computed to establish and preserve the relationship between each respective time series model at each level. The results, based on root mean square percentage error with simulated hierarchical time series data, are promising.
\end{abstract}

\section{Introduction}
Hierarchical time series(HTS) is a well-established technique. Problems requiring HTS are considered complex because of the nature of the time series: the sum of multiple variables forms the corresponding higher level where the relationships among the variables as well as with the hierarchical levels need to be preserved.  
There are a number of approaches available to address univariate or multivariate time series~\cite{RJ2008}. However, it becomes challenging to maintain two-dimensional relationships because of two primary reasons: 
\begin{itemize}
	
	\item At each hierarchical level, individual time series forecasts are required to maintain the relation among the variables of the corresponding hierarchical level.
	\item The forecasts at any hierarchical level should maintain the hierarchical structure over a period of time.
\end{itemize}

Most of the existing approaches apply either a top-down~(\cite{Grunfeld1960, Fogarty1990, Fliedner1999}) or bottom-up approach~(\cite{Orcutt1968, Edwards1969, KinneyJr.1971, Dangerfield1992, Zellner2000}).
 \cite{Athanasopoulos2009}  highlighted that the top-down approach fails to capture the dynamics of individual series in the hierarchy and neglects information based on the trends in the data whereas the bottom-up approach does capture dynamics of individual series and trends. However, the ``noise'' in the bottom levels is aggregated to the top level~(\cite{Athanasopoulos2009}). In an effort to resolve problems with top-down and bottom-up approaches, \cite{Athanasopoulos2009} suggested the top-down approach on the forecasted proportions and applied linear regression to combine the bottom levels to predict the upper level aggregated series~(\cite{Athanasopoulos2009}). 
Inspired by the~\cite{Athanasopoulos2009}'s approach, the proposed approach applies Odds\footnote{Odds of any variable is the likelihood measure representing the proportion of the given variable against the rest of the variables in the given aggregated/summation category.$$Odds(Y_{i=k}) = \frac{Y_{i=k}}{\sum_{i != k}Y_{i!=k}}$$} rather than proportions and solves systems of linear equations rather than using linear regression. 
While Odds preserves the relation among variables horizontally at the given levels, the top-down approach ensures the aggregated sum vertically remains intact. Odds are computed at the given level of hierarchical time series and forecasted. Various systems of linear equations are generated from forecasted Odds and forecasted sum at the upper level while maintaining the relation among the variables to forecast hierarchical values.

\section{Proposed Approach}
The proposed approach is based on Odds,as mentioned earlier. Let's explain the approach with an example - let there be three levels: 
at the given time $t$, $X_{ij}$ be the counts at the bottom-level, $Y_j$ be the mid-level counts and $S$ be the total sum at the top-level sum as shown in Eqs.~\ref{eq:ML}-\ref{eq:HL1}.
\begin{align}
	Y_i =& \sum_{j=1}^{n_j}X_{ij} &\text{Mid-level: $i^{th}$ element} \label{eq:ML}\\
	S =& \sum_{i=1}^m Y_i &\text{Top-level: Total Sum $S$}\label{eq:HL}\\
	 =& \sum_{j=1}^m \sum_{i=1}^{n_j}X_{ij} & \label{eq:HL1}
\end{align}

Since there is only one number at the top-level, it is a univariate time series: simple time series forecasting can be applied to predict the counts at the top level. However, for mid and bottom levels, there exist multiple variables which co-exist with other variables in the given summation. Rather than applying forecasting with univariate time series, odds are computed for variables at the mid-level and bottom-level and forecasted over a period of time.

Since the top-level sum is forecasted, one knows the expected total sum in the future. At middle level, the odds are predicted allowing the correlation with other variables.
Based on odds, linear equations can be easily formulated. At any given point of time, assume we have three predicted odds for three variables: $Y_1$, $Y_2$, and $Y_3$,  defined at middle level are  $\hat{Odds(Y_1)}$, $\hat{Odds(Y_2)}$, and $\hat{Odds(Y_3)}$.

\begin{align}
	\hat{Odds(Y_1)} &= \frac{\hat{Y_1}}{\hat{Y_2}+\hat{Y_3}}
\end{align}
	Adding 1 to both sides
\begin{align}
	1+ \hat{Odds(Y_1)} &= 1 +\frac{\hat{Y_1}}{\hat{Y_2}+\hat{Y_3}} \\
	&= \frac{\hat{Y_1}+\hat{Y_2}+\hat{Y_3}}{\hat{Y_2}+\hat{Y_3}} \label{eq:OddsUL}
\end{align}
	Let $\hat{S}$ be the predicted sum at the upper level as shown in Eq.\ref{eq.:ULS}.
	\begin{align}
			\hat{Y_1}+\hat{Y_2}+\hat{Y_3} = \hat{S} \label{eq.:ULS} 
	\end{align}
Substituting Eq.\ref{eq.:ULS} to Eq.\ref{eq:OddsUL}
\begin{align}
	1+ \hat{Odds(Y_1)} &= \frac{\hat{S}}{\hat{Y_2}+\hat{Y_3}}\\
	\hat{Y_2}+\hat{Y_3} &= \frac{\hat{S}}{1+ \hat{Odds(Y_1)}}
\end{align}
Similarly, the rest of the equations can be formulated.

Since $\hat{S}$ is predicted at the top-level and $Odds$ are predicted for mid-level, the binary matrix based on the linear equations can be designed as below:
\begin{align}
	\begin{bmatrix}
		0 & 1 & 1 \\
		1 & 0 & 1 \\
		1 & 1 & 0
	\end{bmatrix}
	\begin{bmatrix}
		\hat{Y_1} \\
		\hat{Y_2} \\
		\hat{Y_3}
	\end{bmatrix}
	=
	\begin{bmatrix}
		\frac{\hat{S}}{1+ \hat{Odds(Y_1)}} \\
		\frac{\hat{S}}{1+ \hat{Odds(Y_2)}} \\
		\frac{\hat{S}}{1+ \hat{Odds(Y_3)}}
	\end{bmatrix} \label{eq:LinEqs}
\end{align}
The square binary matrix with zeros as diagonal elements and ones as off-diagonal elements with any order, $n$, is always ``invertible''. Eq.~\ref{eq:LinEqs} shows binary matrix of order three. Considering the invertibility of such binary matrix, there would exist a unique solution for the set of linear equations.
\subsection{Problem and resolution}
While there exists a unique solution for the set of linear equations based on the invertible binary matrix, the solution might involve one or two negative values. Since the categorical values are not expected to be negative, this issue is resolved by reducing the negative values to zero and distributing the sum proportionally to the rest of the values of the aggregated hierarchical upper level.

\section{Simulated Study}
\subsection{Data Generation} \label{ssec:DataGen}
\begin{figure}[ht]
	\begin{center}
		\fbox{\includegraphics[width=0.925\columnwidth]{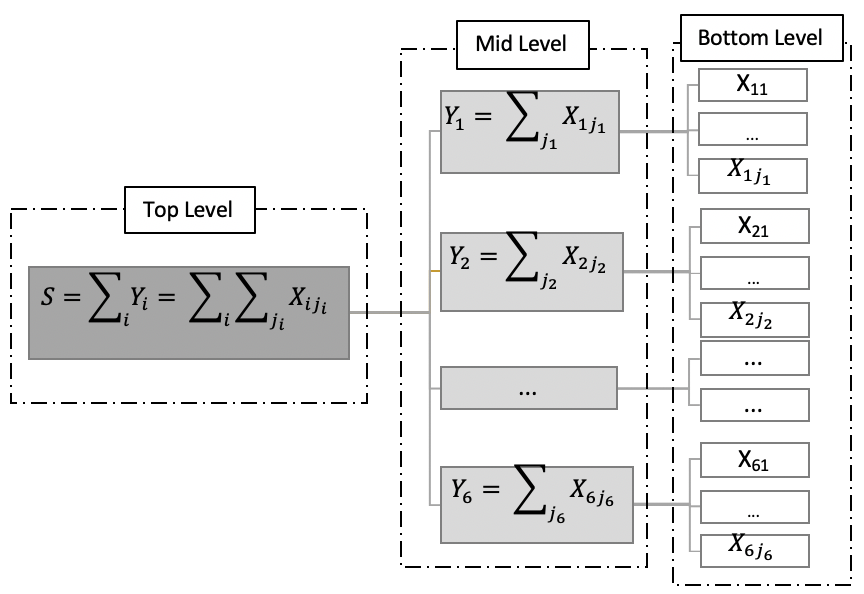}}
		\caption{Hierarchical Time Series Structure}
		\label{fig:hierarchy}
	\end{center}
\end{figure}

The data is simulated for 1000 variables with 1000 time-steps with the INARMA,integer-valued autoregressive (AR) moving-average (MA), approach based on poisson marginals~\cite{Bracher2019}. The simulated data includes only positive integer values. The INARMA approach is modified to introduce variation in AR and MA processes, respectively in variables:
\begin{itemize}
	\item For mid-level: randomly generate six values from one to nine: $j_1$, $j_2$, ..., $j_6$ as shown in Fig~\ref{fig:hierarchy}.
	\item Let $N$ be the sum of the six values generated: $N = j_1+ j_2+ ...+j_6$
	\item For bottom-level: randomly select 'N' variables from 1000 variables generated with the INARMA approach. 
	\item As shown in Fig~\ref{fig:hierarchy}, bottom-level variables build the hierarchy to the top; there is only one variable at the top, whereas there are six at mid-level 
\end{itemize}
\subsection{Experiment}
One hundred different hierarchies were generated using the  approach defined in Section~\ref{ssec:DataGen}. Six different algorithms were applied: ARIMA and five different forms of LSTMs-- Vanila, Stacked, BiDirectional, CNN, and Convolutional were applied to predict the hierarchy over a period of time. A set of 970 time-steps were used to train the model and 30 steps ahead were forecasted, which were compared against actuals to compute root mean square percentage error(RMSPE). One simulation run, start-to-end, takes around 3-4.5 hours on high-performance computing machines(Two Xeon Gold 6230R CPUs, with 4 NVidia V100 GPUs)~\cite{Computing2021}. Considering the run time, it is difficult to optimize hyper-parameters using cross-validation for each run. Since the optimization of hyper-parameters for LSTMs is computationally intensive, they were chosen based on the initial few runs. The results from LSTMs could further be improved with optimized hyper-parameters for each simulation individually.

\section{Results}
\begin{figure}[ht]
	\begin{center}
		\fbox{\includegraphics[width=\columnwidth]{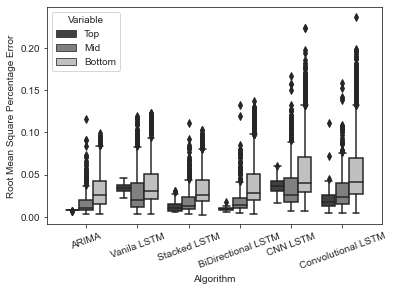}}
		\caption{Percentage mean square error for the simulation study}
		\label{fig:MSPELevels}
	\end{center}
\end{figure}

On comparing against actuals, Figure~\ref{fig:MSPELevels} shows that the ARIMA showed considerable results with 75\% of errors being less than 5\% and the rest of the errors being less than 10\% most of the time. One might debate that ARIMA gave better results as the simulated data is generated with the ARMA approach. But  the hierarchy is designed from bottom to top and only bottom-level variables are generated based on the ARMA approach. On the contrary, stacked LSTM gave similar results.

The RMSPE at top level is even less than 2.5\% with ARIMA, stacked LSTM and BiDirectional LSTM. While stacked LSTM showed little more than 2.5\% beyond 75\% quartile, the other two remained less than 2.5\% for the entire simulation. Although the number of outliers are observed at mid and bottom-levels, these outliers also remain at less than 10\% for ARIMA and stacked LSTM. The results from stacked LSTM can be further improved by appropriately selecting the hyper-parameters. 

\section{Conclusions}
In this paper, we proposed the new approach based on odds forecasting and systems of linear equations to find hierarchical time series. ARIMA and five different versions of LSTM were applied to forecast the value at the top-level and odds at the mid and bottom levels. At mid and bottom levels, systems of linear equations were applied on forecasted odds to forecast the hierarchical time series. The approach was applied on hundred different simulations generated with poisson distribution, and results were compared using mean square percentage error where ARIMA and stacked and bidirectional LSTMs were found to generate results with less than 5\% error. On the contrary, CNN and convolutional LSTMs showed poor results. On the other hand, the errors from stacked and bidirectional LSTMs can further be reduced with appropriate architecture and suitable hyper-parameters. One might consider shifting from one algorithm to another considering the performance at each level. At the end, the results of the simulation study showed favorable results with ARIMA and LSTMs with basic architecture.

\footnotesize
\bibliographystyle{plainnat}
\bibliography{Longitudinal}

\begin{thebibliography}{12}
\providecommand{\natexlab}[1]{#1}
\providecommand{\url}[1]{\texttt{#1}}
\expandafter\ifx\csname urlstyle\endcsname\relax
  \providecommand{\doi}[1]{doi: #1}\else
  \providecommand{\doi}{doi: \begingroup \urlstyle{rm}\Url}\fi

\bibitem[Athanasopoulos et~al.(2009)Athanasopoulos, Ahmed, and
  Hyndman]{Athanasopoulos2009}
George Athanasopoulos, Roman~A. Ahmed, and Rob~J. Hyndman.
\newblock {Hierarchical forecasts for Australian domestic tourism}.
\newblock \emph{International Journal of Forecasting}, 25\penalty0
  (1):\penalty0 146--166, 2009.
\newblock ISSN 01692070.
\newblock \doi{10.1016/j.ijforecast.2008.07.004}.
\newblock URL \url{n}.

\bibitem[Bracher(2019)]{Bracher2019}
Johannes Bracher.
\newblock {A new INARMA(1, 1) model with Poisson marginals}.
\newblock \emph{arXiv}, 2019.
\newblock ISSN 23318422.

\bibitem[Computing(2021)]{Computing2021}
Research Computing.
\newblock {Digital Commons Training Materials}.
\newblock \emph{Kennesaw State University}, 10, 2021.
\newblock URL
  \url{https://research.kennesaw.edu/computing/resources/facilities.php}.

\bibitem[Dangerfield and Morris(1992)]{Dangerfield1992}
Byron~J. Dangerfield and John~S. Morris.
\newblock {Top-down or bottom-up: Aggregate versus disaggregate
  extrapolations}.
\newblock \emph{International Journal of Forecasting}, 8\penalty0 (2):\penalty0
  233--241, 1992.
\newblock ISSN 01692070.
\newblock \doi{10.1016/0169-2070(92)90121-O}.

\bibitem[Dave et~al.(1991)Dave, Fogarty, Blackstone, and Hoffman]{Fogarty1990}
Upendra Dave, D.~W. Fogarty, J.~H. Blackstone, and T.~R. Hoffman.
\newblock \emph{{Production and Inventory Management (2nd Edition)}},
  volume~42.
\newblock 1991.
\newblock \doi{10.2307/2583420}.

\bibitem[Edwards and Orcutt(1969)]{Edwards1969}
John~B. Edwards and Guy~H. Orcutt.
\newblock {Should Aggregation Prior to Estimation be the Rule?}
\newblock \emph{The Review of Economics and Statistics}, 51\penalty0
  (4):\penalty0 409, 1969.
\newblock ISSN 00346535.
\newblock \doi{10.2307/1926432}.

\bibitem[Fliedner(1999)]{Fliedner1999}
Gene Fliedner.
\newblock {An investigation of aggregate variable time series forecast
  strategies with specific subaggregate time series statistical correlation}.
\newblock \emph{Computers and Operations Research}, 26\penalty0
  (10-11):\penalty0 1133--1149, 1999.
\newblock ISSN 03050548.
\newblock \doi{10.1016/S0305-0548(99)00017-9}.

\bibitem[Grunfeld and Griliches(1960)]{Grunfeld1960}
Yehuda Grunfeld and Zvi Griliches.
\newblock {Is Aggregation Necessarily Bad?}
\newblock \emph{The Review of Economics and Statistics}, 42\penalty0
  (1):\penalty0 1, 1960.
\newblock ISSN 00346535.
\newblock \doi{10.2307/1926089}.

\bibitem[Hyndman and Khandakar(2008)]{RJ2008}
RJ~Hyndman and Y~Khandakar.
\newblock {Automatic time series forecasting: the forecast package for R}.
\newblock \emph{Journal of Statistical Software}, 26\penalty0 (3):\penalty0
  1--22, 2008.
\newblock URL \url{http://ideas.repec.org/a/jss/jstsof/27i03.html}.

\bibitem[{Kinney Jr.}(1971)]{KinneyJr.1971}
William~R {Kinney Jr.}
\newblock {Predicting Earnings: Entity versus Subentity Data}.
\newblock \emph{Journal of Accounting Research}, 9\penalty0 (1):\penalty0
  127--136, 1971.
\newblock URL
  \url{http://links.jstor.org/sici?sici=0021-8456{\%}28197121{\%}299{\%}3A1{\%}3C127{\%}3APEEVSD{\%}3E2.0.CO{\%}3B2-8}.

\bibitem[Orcutt et~al.(1968)Orcutt, Watts, and Edwards]{Orcutt1968}
G.H. Orcutt, H.W. Watts, and J.B. Edwards.
\newblock {Data Aggregation and Information Loss}.
\newblock \emph{American Economic Review}, 58\penalty0 (4):\penalty0 773--787,
  1968.

\bibitem[Zellner and Tobias(2000)]{Zellner2000}
Arnold Zellner and Justin Tobias.
\newblock {A note on aggregation, disaggregation and forecasting performance}.
\newblock \emph{Journal of Forecasting}, 19\penalty0 (5):\penalty0 457--465,
  2000.
\newblock ISSN 02776693.
\newblock \doi{10.1002/1099-131x(200009)19:5<457::aid-for761>3.3.co;2-y}.

\end{thebibliography}
\end{document}